# Forecasting Future DDoS Attacks Using Long Short Term Memory (LSTM) Model

Kong Mun Yeen [1], Rafidah Md Noor [1], Wahidah Md Shah [2], Aslinda Hassan [2] and Muhammad Umair Munir [1]

[1] Faculty of Computer Science and Information Technology, Universiti Malaya, Kuala Lumpur, Malaysia
[2] Fakulti Teknologi Maklumat dan Komunikasi (FTMK), Universiti Teknikal Malaysia Melaka (UTeM)

*ABSTRACT*

*This paper forecasts future Distributed Denial-of-Service (DDoS) attacks using deep learning models. Although several studies address forecasting DDoS attacks, they remain relatively limited compared to detection-focused research. By studying the current trends and forecasting based on newer and updated datasets, mitigation plans against the attacks can be planned and formulated. The methodology used in this research work conforms to the Cross Industry Standard Process for Data Mining (CRISP-DM) model. Leveraging cyberattack data from the COVID-19 period (2019–2020), sourced from Digital Attack Map and compiled by Arbor Networks, the study aims to identify recent attack trends and forecast future activity to support proactive mitigation strategies. The dataset was examined using statistical analysis techniques to identify prevailing patterns, with emphasis on the frequency of attacks, the duration of attack instances, and the maximum throughput recorded during each incident. Compared to other deep learning models, the LSTM model is proposed for its ability to learn long-term temporal patterns in evolving DDoS traffic. The performance of LSTM model was evaluated using Mean Squared Error (MSE) under varying neuron counts and window sizes. While the model demonstrated limited predictive accuracy in terms of absolute values, the visual comparison between the predicted and actual data using line charts revealed close alignment in trend patterns. This suggests that the model captures the underlying temporal dynamics of the data, thereby providing a promising foundation for future model optimization and performance enhancement.*

## 1. Introduction

Many cyberattack methods are well known, including but not limited to phishing, spoofing, malware infections, ransomware, and Denial-of-Service (DoS) attacks. A DoS attack occurs when an attacker attempts to disable a service, server, or network. Attackers attempt to make services inaccessible by overwhelming the available resources on the hosting server, infrastructure and/or systems. However, DoS can be easily tracked, as it could contain information about the attacker that can be obtained from network traces and attack logs. Distributed Denial-of-Service (DDoS) is a distributed form of the DoS attack, and it is harder for cybersecurity solutions such as Intrusion Detection Systems (IDS) and Intrusion Prevention Systems (IPS) to detect. This is because the DDoS attacks originate from multiple sources and target one or more victims simultaneously, making it very difficult to pinpoint the real source of the attack.

Government agencies, healthcare providers, and large organizations are among the main targets of cyberattacks.[1]. It is also reported that there is a 372% increase in DDoS attacks on healthcare organizations since the end of 2020. In Germany, the first fatal death due to a cyberattack on a hospital was reported. A higher increase in DDoS attacks was also reported by Cloudflare in their 2021 Q2 security report[2].By understanding the trends of DDoS attack classes, businesses, governments, and organisations can mitigate the incidents by taking preventive and countermeasures to strengthen network security and resiliency towards DDoS targeted attack classes.

## 1.1. Issues and Challenges

DDoS attacks can generally be classified as signature-based and anomaly-based attacks. Most Intrusion Detection Systems can detect signature-based attacks, as there are patterns in the attacks [3]. Anomaly-based attacks, on the other hand, do not have fixed patterns. Hence, it is harder for existing cybersecurity solutions to detect such attacks. DDoS attacks, which are constantly changing and evolving, will be difficult for organizations and other entities to keep their network security in check. The inability to predict attacks leads to poor mitigation planning, which can ultimately impact business continuity. It is estimated that losses due to cybercrimes can reach $10.5 trillion by 2025[2].Most existing research relies on publicly available datasets such as CAIDA (Center for Applied Internet Data Analysis), NSL-KDD, and CICIDS2017 (datasets available from the Canadian Institute for Cybersecurity) [4]. These datasets are outdated and do not reflect the current DDoS attack trends during the COVID-19 pandemic[5]. On top of that, existing detection models proposed by researchers are more focused on protocol and application and network layer attacks, instead of transportation or exploitation-type attacks[3], [6]. This will potentially increase the false negatives when the datasets do not contain new types of DDoS attacks. Many researchers have presented several approaches to detecting and predicting DDoS attacks using various machine learning algorithms. One of the main limitations of machine learning methods is the amount of data they can process and the time needed to process it. This is in comparison to the deep learning approach, which can handle more data[5]. A study from Sahoo et al.[7]shows that using deep learning is proven to be very efficient in predicting DDoS attacks.

## 1.2. Potential Solutions

In the study of DDoS attacks, the Long Short-Term Memory (LSTM) and Recurrent Neural Network (RNN) are some of the techniques used to develop the machine-learning-based detection and prediction models of DDoS attacks. LSTM and RNN belong to the family of deep learning algorithms. This is due to the ability of this family of deep learning algorithms.LSTM is considered one of the most effective techniques for predicting nonlinear, time-variant data compared to other neural network and machine learning methods[3]. This effectiveness is due to the ability of LSTM to learn longer historical data and the ability to solve the gradient problems associated with the Back Propagation RNN technique [8]. LSTM has also been studied for other time-series-based forecasting, such as traffic speeds and the stock market.A predictive system, rather than one limited to detection, would allow the user or organisation to pre-emptively produce a mitigation plan for defending against such attacks.The proposed solution is a process of predicting DDoS attack trends using datasets gathered from 2019-2021 (during the COVID-19 pandemic). Section 2 provides literature reviews of previous research related to this research topic. Section 3 presents the methodology used in the trend study and prediction of DDoS attacks. It explains various tools and methods that have been employed in this research. Section 4reports findings and explains in detail the interpretation of the findings. Finally, section 6 concludes this research and potential future work.

## 2. RELATED WORK

The COVID-19 pandemic has given opportunities to many cyberattacks, targeting various organisations and critical infrastructure across the globe, including but not limited to healthcare services [9]. Many types of attacks were discovered, such as phishing, malware, communications platform compromise (e.g., Zoom, Microsoft Teams), and Denial-of-Service (DoS). DoS and Distributed DoS (DDoS) are very popular types of attack globally due to their being easy to implement but a lot harder to defend against. Due to this, it could be seen as the most dangerous type of cyberattack [5].

In a study from Khan et. al [10], the top ten deadly cyber-security threats were identified during the COVID-19 pandemic. By analysing the range of cyberattack incidents, the study found loose correlations between key events or announcements and cyberattacks. Among the top 10 threats, DDoS is the top-ranked attack as seen by most government and healthcare organisations. The journal concludes that with the rise of ubiquitous computing, there is an increase in cybersecurity threats as well, thus enforcing the need to increase vigilance in defending against them. Figure 1 categorises the DDoS attack types, the detection and mitigation methods. It also highlights prediction tools and techniques, emphasizing deep learning (e.g., LSTM), machine learning, statistical, and knowledge-based approaches.

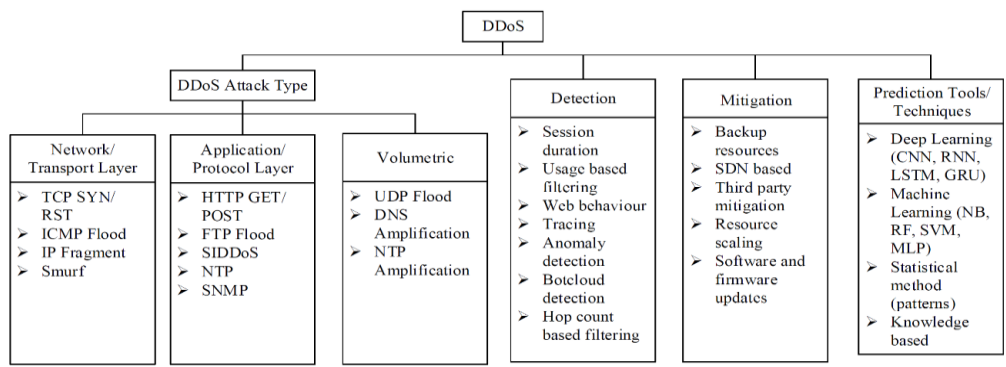

Figure 1. Taxonomy of DDoS

### 2.1. DDoS Prediction Algorithms

The term prediction can be quite misleading and is used interchangeably in the reviewed journals, which led to a lot of confusion. The term prediction could bring different meanings, depending on the context and usage: detecting if a DDoS attack is being deployed, classifying the type of DDoS attack when detected, or forecasting the future trend of DDoS attacks.

In the majority of the journals, the context of prediction refers to predicting if a DDoS attack is happening and classifying the type of attack. In other journals of similar research topics, this is also referred to as DDoS detection. The focus of these journals is the study of features from network measurement logs, such as Wireshark traces, to determine if an attack is happening on the infrastructure and systems.

When a DDoS attack is detected, the next step is to classify the type of DDoS attack to immediately react appropriately against the attack. This task may reuse the same features that are employed by DDoS detection algorithms. The topic of classification is generally discussed in the same journals that study the DDoS detection mechanisms.

A minority of journals researched the topic of predicting DDoS attacks on the premise of determining when the attack will happen in the future, based on time-based historical data of known attacks. In some journals, this context of prediction is also called forecasting. This research project's focus is on this context of prediction.

**2.1.1. Machine Learning Algorithms**

Machine learning algorithms can be generally categorized into three types: supervised, unsupervised, and semi-supervised. In supervised learning, data are labelled and used to train models that classify network traffic according to attack patterns. The model "learns" these patterns from a large dataset and produces a fitted model applicable to similar unseen data[11]. In unsupervised machine learning, data lack labels and are grouped into clusters based on similarities or learned thresholds, while semi-supervised learning refers to both labelled and unlabelled data. However, conventional machine learning often struggles to handle large-scale network traffic effectively. In contrast, deep learning techniques have demonstrated superior performance, producing lower error rates and higher accuracy[12].Researchers have explored multiple deep learning methods to predict DDoS attacks, such as deep multilayer perceptrons (MLP), recurrent neural networks (RNN), long short-term memory (LSTM), convolutional neural networks (CNN), and hybrid combinations, to enhance DDoS detection and forecasting [8], [13]. For example, in [14], Nadeem et al. specifically addressed the challenge of low-rate DDoS (LR-DDoS) detection in Software-Defined Networks (SDN), where traditional approaches often fail due to the stealthy nature of attack traffic. They proposed an RNN-based framework that extracts flow-level features from the CIC DoS 2017 dataset (converted via CICFlowMeter) and deploys the model within an SDN controller for real-time detection. Using Mininet and Ryu controller in their experimental setup, the method achieved 98.59% detection accuracy, which resulted in outperforming conventional classifiers such as Random Forest, SVM, MLP, and CNN. Their results demonstrated that RNNs can effectively capture hidden sequential dependencies in traffic flows that make them highly suitable for detecting low-rate DDoS attacks in programmable network environments.

Similarly, another study [15], showed that RNNs achieved lower error rates compared with Random Forest and demonstrated their ability to capture longer-term historical dependencies, thereby improving detection performance. In addition, Hnamte et al. [16] proposed a-two-stage deep learning model combining LSTM and Autoencoder (LSTM-AE) for network intrusion detection. The authors evaluated this on CICIDS2017 and CSE-CICIDS2018 datasets; the hybrid model showed robust detection capabilities that significantly reduced both false positives and false negatives in dynamic attack scenarios. Some of the main features used in identifying a DDoS attack are the number of packets, the time it takes to send the packets, the packet rate, and the bit rate. Machine learning-based detection has been used by IDS and IPS solutions too[17]. Several machine learning algorithms have been deployed for this purpose, including but not limited to Naïve Bayes, Random Forest, J48 Decision Tree, Support Vector Machine (SVM), Decision Tree, and Multilayer Perceptron (MLP). By using the said models, detection and classification of DDoS attacks can achieve beyond 90% accuracy rate, some even reaching 99%.Support Vector Machine (SVM) and Linear Regression are proposed by Devi S et al. [18]to predict and classify DDoS attacks on Cloud services. To classify DoS and DDoS attacks, SVM is used as it can be kernelized to solve non-linear classification problems. Linear Regression is used for visualizing and forecasting future attacks, due to its ability to predict target variables on a continuous scale. The approach of the forecast is to determine if packets received are sent by an authentic user or by an attacker. It is found that SVM is quite suitable for the classification of UDP-based attacks and somewhat suitable for TCP attacks. But Linear Regression is not suitable for forecasting as it has a high Mean SquaredError of 488.25 for the training dataset and 473.56 for the test dataset.

Research in Distributed Denial-of-Service (DDoS) attack detection has employed various machine learning (ML) and deep learning methods to enhance prediction accuracy. Sahoo et al.[7] evaluated seven ML algorithms, k-Nearest Neighbour (KNN), Naïve Bayes (NB), Support Vector Machine (SVM), Random Forest (RF), Linear Regression, Artificial Neural Network (ANN), and Decision Trees, focusing on detecting Smurf, UDP Flood, and HTTP Flood attacks. Their comparative analysis demonstrated the varying strengths of each algorithm in classifying attack types. In a different approach, Alguliyev et al. [19] proposed predicting DDoS attacks using social media text analysis, specifically Twitter data from the USA. By performing sentiment analysis and employing Convolutional Neural Network (CNN) and an enhanced Long Short-Term Memory (LSTM) model, their method achieved a prediction accuracy of 0.77, highlighting the potential of non-network data sources in proactive attack forecasting. Similarly,[20]explored deep learning approaches for DDoS attacks forecasting, comparing RNN, LSTM, and GRU algorithms, and found that LSTM outperforms other DL and machine learning models in predicting DDoS traffic with high accuracy up to 20 seconds ahead. The lateststudy[21] evaluates multivariate LSTM models for predicting DDoS attacks, comparing their performance with other machine learning models on the CICDDoS2019 dataset, achieving significantly better results than existing techniques in preventing and mitigating DDoS attacks. On the other hand, Messaoud [22]uses LSTM to classify network traffic, which achieves superior performance in capturing temporal dependencies in network traffic.

Deep learning has also been leveraged for more complex pattern recognition. Yuan et al.[15]developed a Bidirectional Recurrent Neural Network (Bi-RNN) model that processed sequences of network traffic traces from large datasets, effectively reducing the error rate from 7.517% to 2.103% compared to conventional ML models. The study suggested future research should explore more diverse DDoS vectors and varied system settings. More recently, Berei et al.[23]used experimental datasets of DoS, DDoS, and normal traffic to evaluate RF, KNN, and SVM models, achieving accuracy levels above 99%. Their use of double feature selection proved effective in mitigating overfitting, reducing model complexity, and improving both training and prediction efficiency. More recently, Afraji et al. [24] examined deep learning-driven defenses for DDoS attacks in cloud environments. The authors categorized the threats into volumetric, protocol, and application-layer types. While CNNs, LSTMs, RNNs, and Autoencoders achieved detection accuracies above 99%, the authors noted key challenges including imbalanced datasets, high computational cost, and the "black-box" nature of DL models.

In addition, Kumar et al. [25] proposed a proactive DDoS detection framework using multiple deep learning architectures, including DNN, CNN, and LSTM, that was trained on the CICDDoS2019 dataset. To improve efficiency, the authors applied Pearson Correlation-based feature selection to remove redundant attributes before training. Their evaluation showed that the DNN model achieved the highest accuracy (98.31%), followed by CNN (97.27%) and LSTM (96.78%). While DNN exhibited the best overall classification performance, LSTM recorded the lowest log-loss (0.45) and high recall, making it well-suited for sequence-based detection scenarios. These findings reinforce that different deep learning architectures provide distinct strengths, with DNN excelling in accuracy, CNN in spatial feature extraction, and LSTM in temporal pattern recognition, highlighting the potential of ensemble or hybrid designs for proactive DDoS defense. Similarly, Saini et al. [26] proposed a synthesized K-fold cross-validation approach to improve the robustness of DDoS detection using multiple ML classifiers. Their framework was tested on several widely used datasets, including CICIDS2017, CICDDoS2019, CSE-CICIDS2018, and NDSec-1, providing a broad evaluation across different traffic conditions. The results showed that Random Forest consistently achieved the highest detection accuracy (up to 99.98%), while other classifiers such as Decision Trees, Logistic Regression, and k-NN also performed well on certain datasets. By combining K-fold validation with diverse ML models, their approach reduced variance, improved generalization, and provided a more reliable benchmark for selecting suitable algorithms in practical DDoS detection scenarios.

## 2.2. DDoS Prevention and Mitigation

Bhardwaj et. al [27] study shows that there is no one-shot comprehensive countermeasure against each known DDoS attack. This increases the complication when cyber attackers keep coming up with new vector threatsand attack derivatives that can avoid detection by IDS and IPS solutions. Hence, they conclude that more research is needed to design and develop effective DDoS prevention and mitigation solutions. Due to the difficulty of detecting and predicting DDoS attacks in advance, at the moment, the ideal time to mitigate DDoS attacks would be at the beginning of the attack execution, preventing it from arriving at the target. This approach is also proposed by Jog et al.[12] by forecasting attacks using the ARIMA model. Other mitigation steps also include keeping up to date with security patches. This helps to prevent exploitation-based attacks that target flaws in the implementation of specific protocols.

Pranggono & Arabo [28]provides a holistic view of how mitigation and prevention of cyber-attacks and DDoS can be approached practically. The preventive steps do not focus on DDoS only, as any weak points can be exploited and contribute to DDoS attacks. User education is very important to bring awareness to how every person and system contributes to a secure infrastructure. The use of Virtual Private Networks (VPN) when working remotely allows a secure working environment, preventing attackers from intercepting communications. Multi-factor authentication (MFA) strengthens account logins to prevent unauthorised access. All software applications and firmware versions should be up to date with the latest security patches to prevent exploitation by attackers.In line with proactive defense, Bitit et al. [29] proposed a forecasting-based solution for DDoS attacks. Their system combines time-series forecasting and online change-point detection to anticipate abrupt shifts in attack traffic volume. By dynamically selecting among multiple statistical models and using change-point alerts to adapt the algorithm models, the approach forecasts attack flow counts in real time. The authors evaluated the proposed algorithm on the CICDDoS2019 dataset, and it significantly outperformed traditional methods like ARIMA and Exponential Smoothing. Their design includes a decision-making module that, based on forecasted attack levels, triggers early alerts so administrators can initiate countermeasures before attacks peak.

More recently, Gilmary et al. [30] developed an intelligent prevention and mitigation system designed for Software-Defined Networking (SDN). Their approach uses a deep neural network (DNN) model trained on real network traffic to quickly tell apart normal and malicious flows. The system first collects and filters traffic features (such as packet size, flow duration, and protocol type) using methods like PCA and Recursive Feature Elimination (RFE). These features are then fed into the DNN, which classifies the traffic in real time. Once an attack is detected, the SDN controller can reroute traffic, update flow rules, or apply rate limits to reduce the attack's impact while keeping legitimate traffic running. Their experiments showed high accuracy (above 95%) and fewer false positives, proving the method is effective for large-scale environments like smart city data centers. However, the authors also pointed out some challenges, including the need for large, labelled datasets, high computational cost, and sensitivity to adversarial attacks. They suggested that future improvements could come from semi-supervised learning and stronger anomaly detection techniques. Similarly, Garba et al. [31] proposed a real-time DDoS detection and mitigation framework for SDN-enabled smart home networks. Their system combines machine learning classifiers (Decision Tree, SVM, KNN, Logistic Regression) with a SNORT intrusion detection system to protect both IoT devices and the SDN controller.

Real-world test bed experiments showed that Decision Trees achieved the best performance with up to 99.5% detection accuracy, while SNORT effectively prevented the controller from being taken offline during TCP SYN flood attacks. The framework also applied feature selection (e.g., PCA) to reduce redundancy and improve efficiency. The authors noted limitations such as dataset size, need for labelled data, and exclusion of deep learning, suggesting future extensions with scalable DL methods. Existing research poses several challenges, including the lack of up-to-date datasets (the latest available is the CICDDoS2019 dataset), specific DDoS attack types, limited feature selection processes (specific DDoS attack types), and the need for scalable ML models. Furthermore, while deep learning has shown promise, its real-world deployment feasibility remains an open question due to computational constraints. This study aims to bridge these gaps by utilizing the dataset during the COVID-19 pandemic to systematically compare real network traffic with the prediction by modelling time-series characteristics of attack volume, duration, and throughput, thereby supporting the development of more resilient and forward-looking cybersecurity solutions.

## 3. METHODOLOGY

To ensure a structured and goal-driven approach, the CRISP-DM (Cross Industry Standard Process for Data Mining) framework is adopted. This methodology provides a comprehensive guide through six key phases—from understanding the problem context and preparing the dataset, to model building, evaluation, and deployment. However, in this article, deployment is excluded.

### 3.1. Dataset Scraping

The dataset is scraped from the Digital Attack Map website (https://www.digitalattackmap.com). The website provides live data visualization of DDoS attacks around the globe. It was built through a collaboration between Google Ideas (now known as Google Jigsaw) and Arbor Networks (now part of Netscout Systems after multiple acquisitions). The data is sourced from more than 330 ISPs in the world. It allows users to explore historic trends and find reports of outages happening on a given day, with rich visualization options and flexible filtering.

To scrape the data, a Chromium-based browser was used (shown in Figure 2). By using the browser's DeveloperTools, the HTTP/HTTPS communications between the browser and the server can be observed. From the Network tab, the file attacks_v2.json is found and separately downloaded from https://www.gstatic.com/ddoz-viz/attacks_v2.json. The size of the file is 163.4 Megabytes.

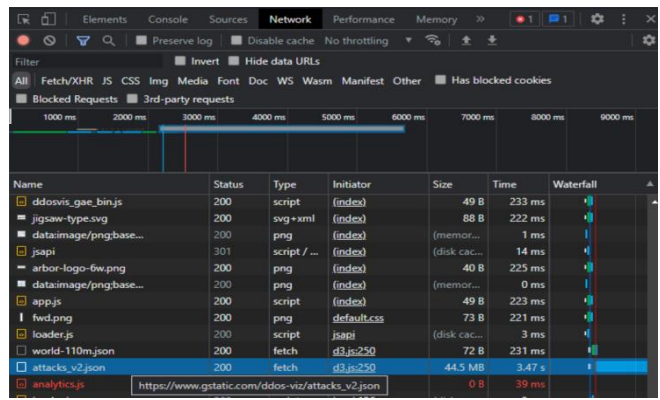

Figure 2. Scraping data using Chromium-based browse

## 3.2. Data Understanding

Since the data is scraped from a source with no documentation, it is critical to explore and understand what is available in the data and how it can be used. Due to the size of the file, common and popular text editors such as Notepad++ cannot be used as the software would crash from the size. So, the file is explored using Python and its relevant libraries. The data was converted into a pandas DataFrame format for easier handling and exploration.

This step aims to understand the dataset in a very general manner, the size, scale, and contents of the data. The dataset consists of 9 features (columns) and 192,525 samples (rows). As there is no proper documentation found from the website on what the columns and values represent, the definitions of the columns are assumed and described in Table 1. Some of these columns will be used to infer other information that is useful for statistical analysis and training of the LSTM prediction model.

Table 1. Dataset column description

| Columnname | Datatype | Assumed column description |
|---|---|---|
| attack_class | STRING | Class of attack type. In this dataset, the values are *Misuse* or *Detector*. |
| dst_cc | []STRING | An array of destination countries of systems under attack. |
| dst_ports | []STRING | An array of destination countries of systems under attack. |
| max_bps | INTEGER | The maximum bit rate that's logged during the attack in bits per second. |
| src_cc | []STRING | An array of source countries of the attacker. |
| src_ports | []INTEGER | An array of source ports of the attacker. |
| start | TIMESTAMP | Start time of the attack in Unix timestamp. |
| stop | TIMESTAMP | Stop time of the attack in Unix timestamp. |
| subclass | STRING | Subclass of the attack type. In this dataset, the values are *Bandwidth, DNS Misuse, ICMP, IP Fragment, Protocol, TCPRST, TCPSYN, Total Traffic,* and *UDP Misuse*. |

For this research project, the study is done on a global basis. With this consideration, the following columns are dropped from the dataset: dst_cc, dst_ports, src_cc, and src_ports. In this dataset, the attacks are divided into the categories as shown in Table 2. Hence, for the rest of this research project, only the attack subclass is used.

Table 2. Attack classes

| AttackClass | AttackSubclass |
|---|---|
| TCPConnection | TCPSYN |
| | TCPRST |
| | TCPACK |
| | Protocol |
| Volumetric | UDPMisuse |
| | ICMP |
| | Bandwidth |
| | TotalTraffic |
| Fragmentation | IPFragment |
| Application | DNSMisuse |

## 3.3. Data Pre-processing

Data pre-processing is an important step in most data analysis activities to provide an appropriate platform for statistical analysis and create an accurate prediction model. The dataset may contain numerous ambiguities, duplicates, errors, and redundant values. Some values are also required to be cleaned or massaged to be used effectively. The approach for data cleansing is to first identify the features and target variables in the dataset. Then quickly review the dataset's value distribution of the features to check for anomalies with the values and decide on the data cleansing and massaging approach. Fortunately, the dataset is already clean with no missing values and imbalances. The next step is to massage the data for more information.

### 3.3.1. Data Massaging

In the previous step, some useful columns werealready identified. Conversions and calculations are performed on the existing columns to enrich the dataset for better statistical analysis and prediction modelling.

**A. Timestamp conversion**

The columns start and stop are in Unix timestamp format. It is converted into a human-readable format and stored inthe columns start time and stoptime with the pandas timestamp format.

**B. Yearly, Monthly, Weekly, and Daily**

To simplify the aggregation of data, columns for years, months, weeks, and dates are created by extracting the information from the column starttime. The aggregated data is only used for visualization in charts and tables. Additionally, daily aggregated data is used as the input dataset to the LSTM model for forecasting future DDoS attack trends. For reporting the statistical values, only the original data set is used. For example, the monthly aggregated data is not derived from daily aggregated data, as this will calculate averages from averaged data and will cause inaccuracies.

**C. Count**

The column count is added with the value 1.0 to simplify the counting of attacks by subclasses and yearly/monthly/weekly/daily periods. When aggregated, the sum is used.

**D. Duration_min**

A new column duration_min is added that contains the attack duration via the formula (Stop – Start)/60, to calculate the attack duration in minutes. When aggregated, the average or mean is calculated.

**E. Max_gbps**

The original column reports this data in bits per second (bps). The values are converted to Gigabits per second (Gbps) for easier interpretation. When aggregated, the average or mean is calculated.

### 3.3.2. Data Aggregation

For statistical analysis and modelling using LSTM, the data is aggregated on a daily, weekly, monthly, and yearly basis. The statistics for the following columns are calculated accordingly:

1. *count* – Sum by *subclass*
2. *duration_min* – Average by *subclass*
3. *max_gbps* – Average by *subclass*

## 3.4. Statistical Analysis and Data Modelling

Statistical analysis was conducted by comparing the values of count, duration_min, and max_gbps on annual, monthly, and weekly bases to determine the general ranking of attack subclasses and assess trend changes before and during the COVID-19 pandemic. Visualizations generated using Matplotlib illustrate how these trends evolved over time. For forecasting, a Long Short-Term Memory (LSTM) deep learning model was employed, as it is well-suited for detecting patterns in sequential data and predicting future trends. The model was developed using TensorFlow with Keras, enabling prediction execution and accuracy evaluation. The parameter settings used for the data modelling is summarised in Table 3.

Table 3 The parameter settings

| Parameter | Value |
| --- | --- |
| Window size | 8, 16, 24 and 32 |
| Neurons | 32, 64 and 128 |
| LSTM Layer | One |
| Optimizer | RMSprop |
| Learning rate | 0.0002 |
| Epoch | 100 |

The network architecture for the model consists of one LSTM layer with 64 neurons. LSTM layer with 32, 64 and 128 neurons were used. But 64 neurons seem to provide the best balance between accuracy and computation performance. Additional layers were also tested, but at rough evaluation, it did not provide any clear advantages over a single layer LSTM model.

## 3.5. Training & Scoring

### 3.5.1. Splitting Data and Normalisation

The dataset is split into sizes of 50% for training, 20% for validation, and the remaining 30% for testing. Training data and validation data are used for fitting the LSTM model. The testing data is used to evaluate the LSTM model.

The count, duration, and maximum throughput are also normalized by dividing the values by the standard deviation of the data array. The equation for normalization is shown below.

$$\sigma = \sqrt{\frac{\sum_{i-1}^{N}(x_i - \mu)^2}{N-1}} \qquad (1)$$

The dataset is used to create TensorFlow datasets of selected window size, with a target value as shown in Figure 3. The window size determines the number of input values in a time series dataset to be used by the LSTM model to predict the next output value.

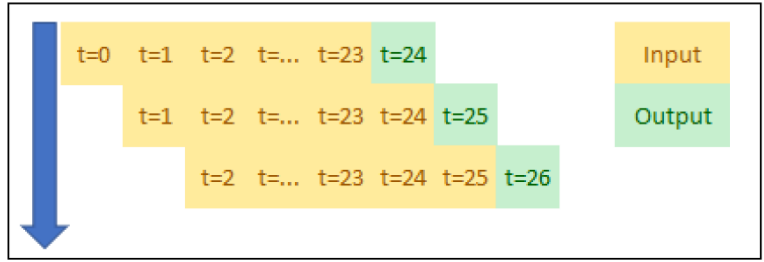

Figure 3. TensorFlow dataset of Window size = 24

### 3.5.2. Evaluation Metric

To evaluate the LSTM model, MeanSquared Error (MSE) and Mean Absolute Error (MAE) are used [5].MSE is the average squared error of the difference between the target and the predicted value by the LSTM algorithms. In MSE, small errors will be magnified as the errorsare squared to overestimate how bad the model is. A smaller MSE value means the prediction is accurate with less error [18].

$$MSE = \frac{1}{N}\sum_{j=1}^{N}(yj - \hat{y}j^2) \quad (2)$$

Where $y$ represents the target value, $\hat{y}$ represents the predicted value from the LSTM model, and N represents the number of data points.

## 4. RESULTS AND DISCUSSION

### 4.1. Data Exploration & Analysis

The dataset, spanning from 1 January 2015 to 1 May 2021, contains detailed records of 192,525 DDoS attacks. Collectively, these attacks account for a total duration of 825,549,256 seconds— equivalent to approximately 628.27 years—an exceptionally high figure caused by multiple overlapping attacks occurring simultaneously against one or more victims. The highest recorded attack throughput reached 1.46 Tbps, while the longest single attack lasted 604,944 seconds (7 days).

When comparing the distribution of attack duration and attack throughput between years 2019 and 2020, there is a general increase year-on-year. There is a 63.65% increase in attack duration between 15 and 30 minutes and a 74.13% increase in attack duration between 1 hour and 1 day. Attack throughput also increased, reaching a 38.12% increase for throughput between 10 Gbps and 100 100Gbps, and a 59.59% increase for throughput between 100 Gbps and 1Tbps. It is not so obvious in Figure 4 due to the size, but the number of attacks reaching beyond 1 Tbps has grown from 277 to 538 incidents, which is a 94.22% increase.

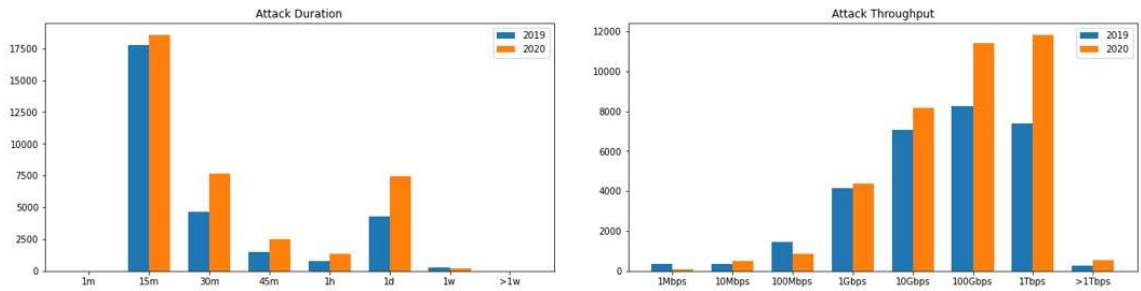

Figure 4. Attack Duration and Attack Throughput Distribution

The count of each attack subclass in the entire database is shown in Table 3. The all-time top 3 attacks are Total Traffic, UDP Misuse, and IP Fragment.

Table 3: Attack counts

| Attack subclass | Count |
|---|---|
| Total Traffic | 62589 |
| UDP Misuse | 50201 |
| IP Fragment | 30822 |
| TCP SYN | 17585 |
| ICMP | 10742 |
| Bandwidth | 7121 |
| Protocol | 7058 |
| TCP RST | 4511 |
| DNS Misuse | 1896 |

The analysis was then extended to examine each attack subclass individually, with the aim of identifying specific characteristics that influenced overall attack statistics and detecting shifts in attacker strategies. By breaking down DDoS trends into subclass categories, it becomes possible to determine which types were most prevalent over time, particularly during the COVID-19 pandemic years of 2019 and 2020.

As shown in Figure 5, the top three subclasses in both years—Total Traffic, UDP Misuse, and IP Fragment, remained consistent in terms of attack counts. However, UDP Misuse and IP fragmentation recorded substantial growth between 2019 and 2020. Notably, while ICMP attacks did not appear among the top three subclasses, they exhibited the highest year-on-year growth rate, highlighting a rapidly increasing trend that could signal evolving attack tactics.

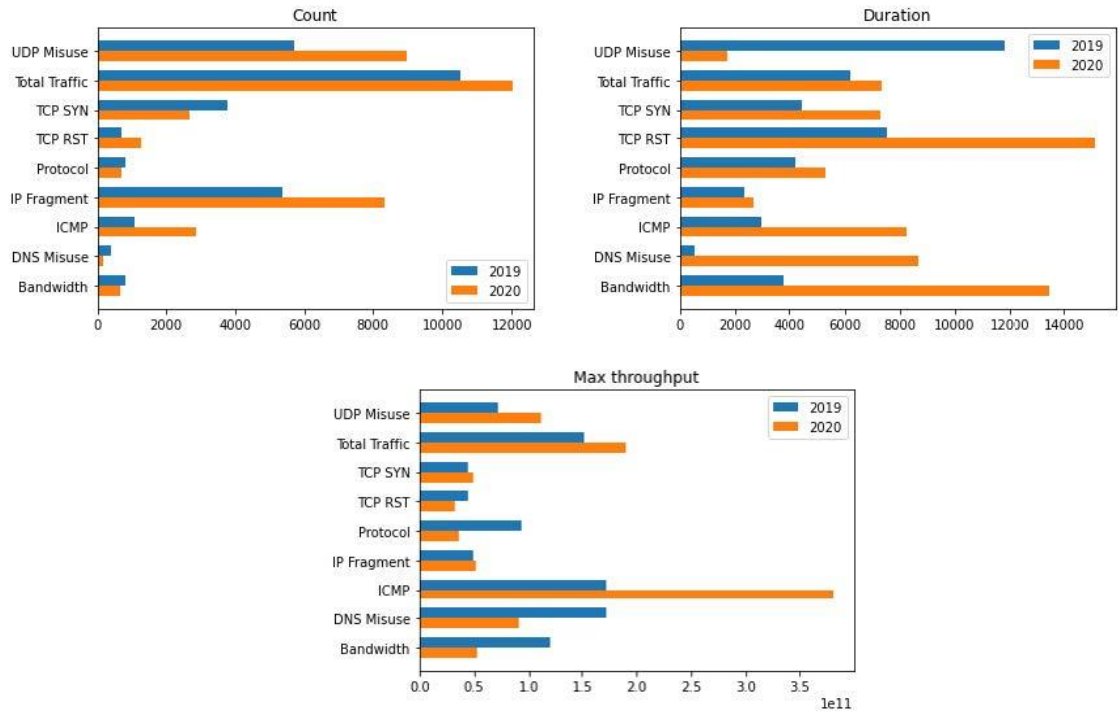

Figure 5. 2019 vs 2020 statistics

### 4.2. Prediction Results

Based on the data exploration and analysis of DDoS attacks from 1 January 2015 to 1 May 2021, a prediction model was developed to forecast the count, duration_min, and max_gbps of various DDoS attack subclasses. The dataset contains 192,525 recorded attacks, with a combined attack duration of 825,549,256 seconds (approximately 628.27 years), an unusually high figure due to overlapping attacks targeting one or more victims simultaneously. The maximum observed throughput reached 1.46 Tbps, and the longest single attack lasted 604,944 seconds (7 days).

Figure 6,7 and 8 show the predicted vs actual pattern for the three parameters. Visually, by using the above parameters for window size, neurons and epochs, the model can forecast the trend for the count and max_gbps. Spikes in the forecasting trend were predicted with reasonable accuracy, even though the values did not fully match. But for duration_min, the trend is comparatively flatter and unable to detect the spikes, possibly due to higher differences between each timestep value.

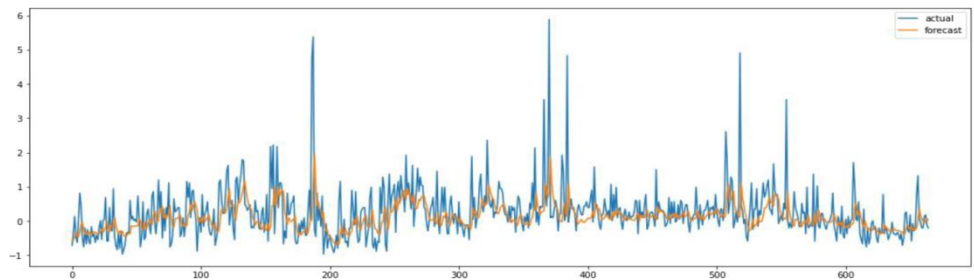

Figure 6. Predicted vs actual count of Total Traffic

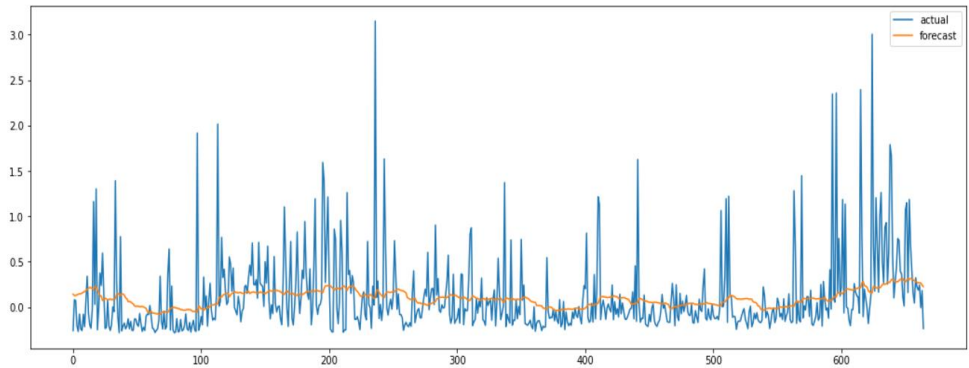

Figure 7. Predicted vs actual attack duration of Total Traffic

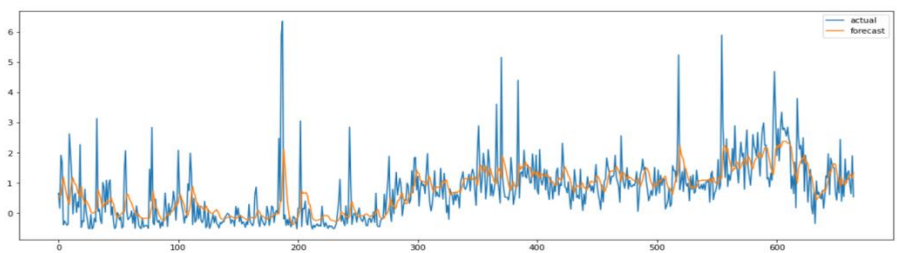

Figure 8. Predicted vs actual maximum throughput of Total Traffic

The model is evaluated by cycling through the values for window sizes and neurons. The ideal values are found to be window sizes of 24 and 64 neurons, giving the best balance between accuracy and computation resources. For the selected parameters, the training Mean Square Error is 1.12, and the test Mean Square Error is 0.52. However, based on the error values, the model performance was not highly accurate. But based on the results in Table 4, the model is seen to be learning very slowly since the error rate is getting lower by increasing the window size and neurons.

Table 4.Model evaluation

|  |  | Neuron | | | | | |
|---|---|---|---|---|---|---|---|
|  |  | 32 | | 64 | | 128 | |
|  |  | Train MSE | Test MSE | Train MSE | Test MSE | Train MSE | Test MSE |
| Window size | 8 | 1.1293 | 0.5278 | 1.1293 | 0.5429 | 1.1293 | 0.5411 |
|  | 16 | 1.1263 | 0.5286 | 1.1263 | 0.5149 | 1.1263 | 0.5136 |
|  | 24 | 1.1198 | 0.5298 | 1.1198 | 0.5158 | 1.1198 | 0.5167 |
|  | 32 | 1.1226 | 0.5266 | 1.1226 | 0.5132 | 1.1226 | 0.5363 |

The statistical analysis in this research examined global attack subclasses by evaluating metrics such as attack volume, time, duration, and affected ports, similar to the approach in D. Kwon et al. [32]. By trending attacks across count, duration, and throughput, correlations between these features were explored to understand execution patterns and potential impacts. For instance, simultaneous increases in count, duration, and maximum throughput suggest heightened attacker aggressiveness, as observed in Total Traffic, IP Fragment, and ICMP attacks. Conversely, when attack counts rise but durations fall, it indicates short, burst-like activity, such as in UDP Misuse attacks. Notably, no attack type displayed a consistent downward trend across all metrics, highlighting the need for continuous vigilance against both long-standing and emerging threats. Studying these patterns at varying granularities provided deeper insights into attacker behaviors and evolving strategies.

For prediction, a Long Short-Term Memory (LSTM) model was implemented using a window size of 24, a single LSTM layer with 64 neurons, and 100 training epochs to forecast attack metrics. While the model exhibited relatively high error rates in evaluation, visual inspection of predicted trends against actual data revealed promising alignment, suggesting that the approach is on the right track. The model captured unusual spikes in daily attack trends with reasonable accuracy, although predicted spike magnitudes were often up to 50% lower than actual values. Despite this gap, the ability to anticipate significant surges in attack activity is valuable for proactive defense measures. These findings indicate that further refinements—such as tuning hyperparameters, increasing dataset diversity, or incorporating additional features—could enhance forecasting accuracy. Overall, integrating statistical analysis with LSTM-based predictions provides both retrospective and forward-looking perspectives, enabling organizations to better understand attack dynamics and prepare for evolving cyber threats.

## 4.3. Threat Landscape: Attacker Strategies and the Need for Prediction

The evolution of Distributed Denial-of-Service (DDoS) attacks reflects a continuous arms race between attackers and defenders. Today's campaigns are no longer limited to simple volumetric floods; instead, they combine multiple vectors such as total traffic floods, UDP misuse bursts, IP fragmentation, and ICMP flooding, which are executed in varying durations and intensities. Statistical analysis from this research underscores these dynamics, where the simultaneous increases in attack count, duration, and throughput indicate periods of heightened aggressiveness, while sudden bursts of high-frequency but short-lived activity, such as in UDP misuse attacks, reveal attackers' preference for unpredictable, hard-to-detect disruptions. Notably, the absence of any consistent downward trend across subclasses highlights that attackers are not abandoning older methods but rather refining and combining them with newer strategies, making the landscape both persistent and adaptive.

This adaptability is further amplified by the rise of IoT-driven botnets and "DDoS-for-hire" services, which enable attackers to mobilize diverse attack patterns at scale and on demand. Such diversity in attack execution was reflected in the dataset, where no subclass remained stable; each displayed shifting peaks in volume, time, or duration. These findings mirror global observations that attackers increasingly tailor their methods to bypass static defenses, alternating between long-duration floods to exhaust resources and evasive spikes to probe weaknesses.

Against this backdrop, reactive defense is insufficient. By the time anomaly-based detection flags an active flood, damage in the form of downtime, service degradation, or reputational loss may already be done. Predictive models such as Long Short-Term Memory (LSTM) networks directly address this gap. Despite exhibiting some errors in magnitude forecasting, the implemented LSTM was able to capture unusual spikes and anticipate surges in daily attack activity. Even partial foresight into impending attack waves enables pre-emptive resource allocation and the initiation of countermeasures, making predictive analytics far more valuable than purely reactive approaches. The integration of statistical trend analysis with predictive forecasting, therefore, offers advantages, i.e., understanding historical attacker behaviors while proactively preparing for their next move.

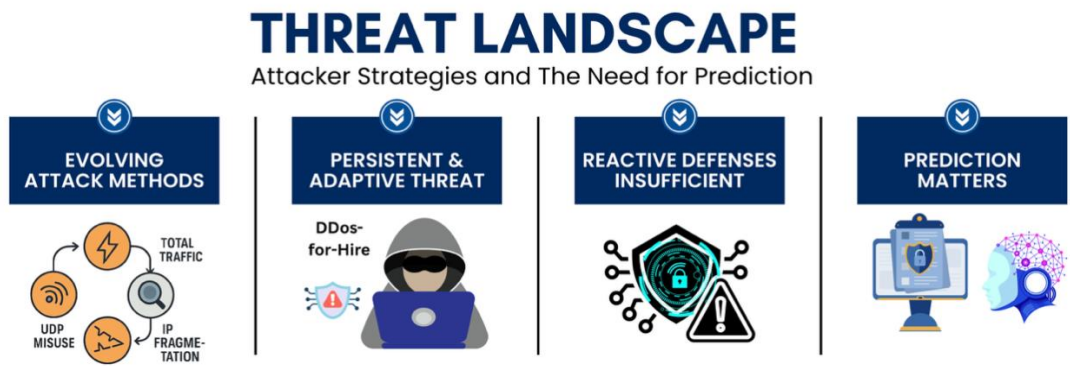

Figure 9. The Evolving Threat Landscape of DDoS Attacks

## 5. CONCLUSIONS

The LSTM model appears to be an effective approach for predictingattack trends, even though the model evaluation is not very conclusive. This is based on how close the predicted trend line is compared to the actual data.In future research, it is proposed to use an updated data set that contains a better classification of attacks and more dimensions. This allows for more comprehensive statistical analysis per dimension, for example, by industry. New attack types may have also been discovered and classified accordingly. The proposed LSTM model could also be enhanced further to improve the prediction of irregular spikes by adjusting more hyperparameters available to the model or by adding more layers to the neural network. Additionally, a forecasting model that utilizes more features to predict the next timestep value can be developed.

## ACKNOWLEDGEMENTS

The authors would like to thank the Center for Advanced Computing Technology (C-ACT), Fakulti Teknologi Maklumat dan Komunikasi (FTMK), Universiti Teknikal Malaysia Melaka.

## AUTHORS

**KONG MUN YEEN** is a seasoned telecommunications professional and Senior Consultant at Orbitage, with extensive experience delivering specialized training across Malaysia, Indonesia and many more. She has developed and conducted courses such as Certified Internet Protocol Associate (CIPA), Certified Internet Protocol Engineer (CIPE), GSM & GPRS Overview, 3G Overview, and LTE Planning, Signalling, and Optimization. Previously a Senior Trainer at ULearn, she holds a Bachelor's Degree in Electrical and Electronics Engineering from Universiti Tenaga Nasional (2003) and a Master of Data Science from Universiti Malaya (2022).

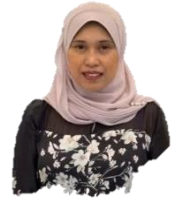

**RAFIDAH MD NOOR** received the BIT degreefrom Universiti Utara Malaysia, in 1998, theM.Sc. degree in computer science from Universiti Teknologi Malaysia, in 2000, and the Ph.D.degree in Computer Science from Lancaster University,U.K., in 2010. She is currently a Professor at theDepartment of Computer Systems and Technology, Faculty of Computer Science and InformationTechnology, Universiti Malaya. Her research is related to the field oftransportation systems in the computer science research domain, includingvehicular networks, wireless networks, network mobility, quality of service,and the Internet of Things.

**WAHIDAH MD SHAH** holds her Bachelor of Information Technology from Universiti Utara Malaysia, Master of Computer Science from Universiti Teknologi Malaysia and PhD in Computer Science from Lancaster University, UK. She is currently a Senior Lecturer in the Department of Computer System and Communication at Universiti Teknikal Malaysia Melaka. She is a member of the Information Security, Digital Forensic, and Computer Networking research group. Her research interests include system and networking, wireless ad-hoc networking, cyber-physical systems (CPS) and IoT related technology.

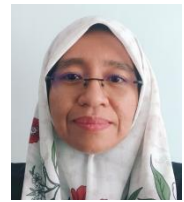

**ASLINDA HASSAN** received her PhD degree in Electrical Engineering, from Memorial University of Newfoundland, St. John's, NL, Canada in 2014. She received M.Sc. degree in Computer Science, from Universiti Teknologi Malaysia (UTM) and B.Sc. degree in Business Administration with honors, from University of Pittsburgh, Pittsburgh, PA, USA in 2001 and 1999, respectively. In 2004, she joined Universiti Teknikal Malaysia Melaka, where she is currently a Senior Lecturer at Faculty of Information and Communication Technology. Her research interests include in vehicular ad hoc network, vehicular communication, wireless ad-hoc network, wireless sensor network, wireless communication, ad hoc routing protocols, cyber-physical systems (CPS), Internet of Things (IoT), network performance modelling and analysis as well as network programming interfaces.

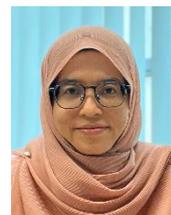

**MUHAMMAD UMAIR MUNIR** received the B.S.C.S. degree (Hons.) from the University of Central Punjab, in 2017, and the master's degree in computer science from Universiti Malaya. He is currently a PhD candidate at Universiti Malaya and Research assistant to Professor Rafidah Md Noor and Associate Professor Dr. Ismail Ahmedy. His research interests include the Internet of Things, wireless sensor networks, vehicular communication, and software quality.

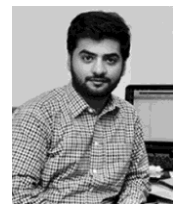